\documentclass[11pt,twoside]{article}

\usepackage{asp2006}
\usepackage{epsf}

\begin{document}
\title{Astronomical Applications for ``Radial Polarimetry''} 
\author{Frans Snik}  
\affil{Sterrekundig Instituut Utrecht, Princetonplein 5, 3584 CC  Utrecht, the Netherlands\\
\texttt{f.snik@astro.uu.nl} }

\begin{abstract} 
Many objects on the sky exhibit a centrosymmetric polarization pattern, particularly in cases involving single scattering around a central source. Utilizing a novel liquid crystal device (the ``theta cell'') that transforms the coordinate system of linear polarization in an image plane from Cartesian to polar, the observation of centrosymmetric polarization patterns can be improved: instead of measuring Stokes $Q$ and $U$ on the sky, one only needs to measure Stokes $Q'$ in the new instrument coordinate system. This reduces the effective exposure time by a factor of two and simplifies the polarization modulator design. According to the manufacturer's specifications and to measurements in the lab, the liquid crystal device can be applied in the visible and NIR wavelength range. Astronomical science cases for a``radial polarimeter'' include exoplanet detection, imaging of circumstellar disks, reflection nebulae and light echos, characterization of planetary atmospheres and diagnostics of the solar K-corona. The first astronomical instrument that utilizes a theta cell for radial polarimetry is the S$^5$T (Small Synoptic Second Solar Spectrum Telescope), which accurately measures scattering polarization signals near the limb of the sun. These observations are crucial for understanding the nature and origin of weak, turbulent magnetic fields in the solar photosphere and elsewhere in the universe. A ``radial polarimeter'' observing a slightly defocused point source performs one-shot full linear polarimetry. With a theta cell in a pupil plane, a beam's linear polarization properties (e.g. for calibration purposes) can be fully controlled through pupil masking.
\end{abstract}

\section{The Theta Cell}
A classical polarimeter measures (a part of) the Stokes vector $S=(I,Q,U,V)^T$ for which the coordinate system is uniform on the sky. However, many sources on the sky have a distinct centrosymmetric pattern of linear polarization, particularly in cases of single scattering around a central source. By applying a transformation from a Cartesian to a polar coordinate system, this linear polarization pattern can be described by a single Stokes parameter $Q'$:
\begin{eqnarray}
Q' & =  & Q\cos(2\phi)+U\sin(2\phi)\\
U' & =  & -Q\sin(2\phi)+U\cos(2\phi)\textrm{,}
\end{eqnarray}
with $\phi$ being the azimuth angle as measured from the center of the centrosymmetric pattern. We define positive Stokes $Q'$ in the instrument frame to correspond to an azimuthally oriented polarization pattern on the sky.\\
\newpage
Such a polarization coordinate transformation can be achieved optically with a so-called ``theta cell'' \citep{thetacell}\footnote{See also \texttt{www.arcoptix.com/radial\_polarization\_converter.htm}.} positioned in an image plane. This theta cell is composed of twisted nematic liquid crystal molecules that locally rotate the direction of linear polarization cf. Equations 1 and 2. All liquid crystal molecule chains are aligned on one side of the device by a linear rubbing on a substrate (the direction of which determines the direction of Stokes $Q'$) and are made to twist by a circular rubbing on the second substrate. This liquid crystal device is achromatic, because it relies on waveguiding of the incident light to rotate its linear polarization direction and not on retardance like a half-wave plate does. Indeed, our lab measurements show that in the wavelength range of 400--650 nm the rotation of linear polarization is accurate to within $\pm2^\circ$ over the entire clear aperture. No degradation towards the blue is found, which would indicate a break-down of the waveguiding condition. The manufacturer quotes an operating wavelength range up to 1.7 $\mu$m and a temperature range up to 50$^\circ$C. The device does, in principle, not modify circular polarization (Stokes $V$), although the twisted molecules might exhibit circular dichroism to some degree. But this has not yet been investigated.\\
The major advantage of using a theta cell is that linear polarimetry of a centrosymmetric source is no longer composed of Stokes $Q$ and $U$ measurements, but a single Stokes $Q'$ measurement suffices. This dramatically simplifies the polarimetric modulator design and reduces the required observing time by a factor of two. Averaging over a radial polarization pattern is then furnished by averaging over Stokes $Q'$. Furthermore, spurious polarization signals due to misalignment between the individual $Q$ and $U$ observations \citep[e.g.][]{misalignment} are avoided. Astronomical ``radial polarimetry'' is ideally performed with a theta cell in prime focus of the telescope to minimize instrumental polarization. Because the theta cell works both ways, transforming radial polarization patterns into a constant linear pattern as well as the other way around, constant instrumental polarization shows up as a ``butterfly pattern'' in the Stokes $Q'$ measurements. In contrast to other polarization optics, the alignment of a theta cell with respect to the optical axis is critical in the sense that the central source in the image plane needs to be accurately aligned with the center of the circular pattern of the theta cell.

\section{Science Cases}
Several science cases in different areas of astronomy could potentially benefit from radial polarimetry.

\subsection{Imaging of Circumstellar Scattering Structures}
Any kind of circumstellar matter that exhibits single scattering of light from a central star is observed to have a centrosymmetric, azimuthal polarization pattern with respect to that central source. In fact, such patterns are used to identify the position of the source when it is obscured by an optically thick cloud. Scattering circumstellar material is found in the form of clouds around young stellar objects \citep{YSO}, protoplanetary and debris disks \citep{disks} and (planetary) nebulae \citep{nebulae}. Particularly in cases of faint or close-in disks or nebulae, linear polarimetry is a powerful differential technique to obtain images of the circumstellar structure. Radial polarimetry around bright sources could therefore improve on the imaging of the circumstellar matter.  Another application is to accurately measure light echos \citep{lightecho} and thereby establish the distance to the central object. Also in the case of active galaxies, the material around the nucleus is bright due to scattering and exhibits a centrosymmetric polarization pattern \citep{activegalaxies}. \\
Several mechanisms are known to modify the pure centrosymmetric pattern due to single scattering: multiple scattering, and magnetic fields by aligning the scattering dust grains \citep{alignedgrains} and through the Hanle effect \citep{Nordsieck}. It is interesting to remark that the small deviations of centrosymmetry are all contained in a Stokes $U'$ measurement of a radial polarimeter. Large deviations would also end up in $-Q'$. This means that if one is primarily interested in e.g. imaging the multiple scattering disk inside a reflection nebula \citep{multiplescatteringdisk} or in measuring the weak circumstellar magnetic fields, one could focus on the Stokes $U'$ observations with a radial polarimeter.

\subsection{Exoplanet Detection}
A specific and exciting example of a circumstellar scattering object is an exo\-planet's atmosphere and/or surface. Generally, exoplanets are substantially linearly polarized perpendicular to the scattering plane. However, the scattered light off an exoplanet is about 10$^8$--10$^{10}$ times fainter than the (unpolarized) light of the central star, which completely outshines the planet. Sensitive polarimetry is therefore a promising differential technique for directly imaging and characterizing exoplanets. Unfortunately, in most cases it is not known \textit{a priori} where the possible exoplanet is located with respect to its host star, so searches have to be performed in both Stokes $Q$ and $U$. Since it is expected that total exposure times of several hours are required to obtain sufficient polarimetric sensitivity, it pays off to reduce the observation dimensionality by implementing a theta cell and only operating the instrument in Stokes $Q'$. The theta cell is ideally combined with a coronographic mask since both are located in an image plane. If the theta cell is positioned after the AO, it only introduces flat field variations.

\subsection{Characterization of Multiply Scattering Planetary Atmospheres}
As described by \citet{neptune}, the limb polarization of Neptune and Uranus in opposition exhibit a centrosymmetric polarization pattern. In this case, this is due to multiple scattering. Radial polarimetry may be applied in order to measure the average degree of linear polarization and obtain spectra of the scattered light in the atmosphere to compare with radiative transfer models.

\subsection{Solar Atmosphere Diagnostics}
The solar K-corona is created by Thomson scattering, and therefore it is azimuthally polarized. Using a theta cell and a Stokes $Q'$ polarimeter, the faint corona could be more rapidly imaged than with conventional coronographic instruments. Also many chromospheric and high photospheric lines are linearly polarized parallel to the solar limb due to scattering. The linearly polarized spectrum measured close to the limb is called the Second Solar Spectrum \citep{sss}, because of the apparent lack of correlation of the line polarization with the intensity spectrum. This line polarization together with the Hanle effect yields a unique way of measuring weak, turbulent magnetic fields in the solar atmosphere. This is the science goal of the first radial polarimeter, the S$^5$T \citep[the Small Synoptic Second Solar Spectrum Telescope; ][]{S5T}. The typical polarization signatures to be measured are of the order 10$^{-3}$--10$^{-5}$. Because integration of all the light along the solar limb leads to polarization cancellation, the Second Solar Spectrum is usually measured with a large-aperture solar telescope to obtain a sufficient photon flux. Using a theta cell and measuring the average $Q'/I$ signal along the limb, this large aperture requirement is dropped, and the S$^5$T is indeed designed to reach a polarimetric sensitivity of 10$^{-5}$ with an aperture of only 5 cm. The prototype already obtained a Second Solar Spectrum with a spectrum resolution of 0.5 \AA~and a polarimetric sensitivity of 6$\cdot$10$^{-5}$ within four minutes of total exposure \citep{S5T}. The complete instrument is currently being designed and constructed to perform automated daily measurements of the Second Solar Spectrum for a duration of eleven years in order to diagnose the relation of the weak turbulent magnetic fields in the solar photosphere with the solar dynamo. The absence of an eleven-year variation of the Second Solar Spectrum could prove the existence of local dynamo action by the granulation, which has implications for the understanding of turbulent magnetic field generation in many astrophysical contexts.

\section{Linear Polarimetry of a Defocused Point Source}
A different trick can be performed with a theta cell in an image plane that enables one-shot full linear polarimetry of a (bright) point source. By defocusing the instrument, an extended spot is obtained that has a butterfly pattern in linear polarization. The spot's polarization pattern is described in polar coordinates $(r,\phi)$ by:
\begin{equation}
\frac{Q'}{I}(r,\phi) = P\cdot\cos\big[2\cdot(\phi+\theta)\big]\textrm{,}
\end{equation}
with
\begin{equation}
P = \frac{\sqrt{Q^2+U^2}}{I}\textrm{, }
\theta = \frac{1}{2}\arctan\Bigg(\frac{U}{Q}\Bigg)\textrm{.}
\end{equation}
Note that polarization modulation for Stokes $Q'$ is not necessary, since the intensity pattern after the theta cell and a polarizer is ideally described by
\begin{equation}
I(r,\phi) = I_0(r) \cdot \Big[\frac{1}{2}+\frac{1}{2}P\cdot\cos\big[2\cdot(\phi+\theta)\big] \Big]\textrm{.}
\end{equation}
With this set-up, complete linear polarimetry is performed with a single measurement, which renders it ideal for time-resolved measurements of e.g. GRB afterglow polarization and perhaps even pulsar properties. Note that by applying curve-fits according to Equations 3 and 5, the linear polarization measurement is much less sensitive to noise than in the case of classical modulation techniques. Of course the photon flux per detector pixel is reduced by defocusing, so this method is only suitable for sources that are not photon starved.

\section{Complete Linear Polarization Calibration}
The properties of the theta cell can also be utilized for calibration purposes. Positioning a theta cell concentrically on a uniform circular pupil image reduces the linear polarization degree in the image plane to zero, independent of the input polarization. This furnishes an accurate method to determine an instrument's zero-level of linear polarization. Especially for spectropolarimetry no useful depolarizers are available since most depolarizers use spectrally dependent polarization modification to yield broad-band unpolarized light. By applying a ``pizza-slice'' masking to the theta cell aperture in a pupil plane and feeding it with fully polarized light, also any non-zero linear polarization can be created to calibrate the instrument's polarization scale, which is particularly useful when it is designed to measure very low fractional polarization. For a transmitting slice with an angular aperture $\Delta\psi$, the emergent degree of linear polarization in the image plane is given by:
\begin{equation}
P=\Bigg|\frac{\sin(\Delta\psi)}{\Delta\psi}\Bigg|\textrm{, }0<\Delta\psi\leq2\pi\textrm{.}
\end{equation}
Obviously light levels change when varying $\Delta\psi$. The angle of linear polarization $\theta$ is determined by the angle of the input polarization and by the angular position of the slice. If the input is $-Q$, then the output angle is identical to the intermediate angle of the aperture slice.

\acknowledgements F.S. acknowledges travel funds to the Astropol 2008 conference from LKBF.

\end{document}